\documentclass[final,5p,times,english,twocolumn]{elsarticle}

\usepackage{amssymb}
\usepackage{tensor}

\usepackage{amsmath}
\usepackage{xcolor}
\usepackage{subcaption}
\usepackage{dsfont}
\usepackage{mhchem}
\usepackage{background}

\journal{arXiv}

\usepackage{hyperref}
\hypersetup{
     colorlinks=true,
      linkcolor=blue,
}

\DeclareMathOperator{\arsinh}{arsinh}
\DeclareMathOperator{\erf}{erf}

\backgroundsetup{contents=}
\begin{document}

\begin{frontmatter}



\title{High current ionic flows via ultra-fast lasers for fusion applications}


\author{Hartmut Ruhl and Georg Korn}

\address{Marvel Fusion, Theresienh\"ohe 12, 80339 Munich, Germany}

\begin{abstract}
In the present paper we introduce a new accelerator
concept for ions. The accelerator is nano-structured
and can consist of a range of materials. It is capable of generating
large ionic currents at moderate ion energies. The nano-structures can
be tailored towards the accelerator thus being capable of driving ion
beams with very high efficiency. The accelerator is powered by laser
arrays consisting of many repetitive and efficient lasers in
the $100 \, \text{J}$ range with ultra-short intense laser
pulses. Combining nano-structures and the proposed ultra-short pulse
lasers can lead to new levels of spatio-temporal control and energy
efficiency for fusion applications.
\end{abstract}


\end{frontmatter}


\tableofcontents

\section{Introduction}\label{intro}
The indirect drive ICF approach has recently achieved an impressive milestone
at NIF \cite{banks2021significant,Zylstra2022,
  Abu-Shawareb2022,Kritcher2022,Zylstra2022b,DOE2022,
  LLNL2022,NYT2022,Science2022} demonstrating the principal viability
of inertial confinement fusion for energy production. A detailed
overview over ICF is given by \cite{atzeni2004physics}.

The coupling efficiency of an external energy source to fusion fuels is a
challenging problem. It  is known that nano-structures can tailor and
greatly enhance the coupling between lasers and matter via a range of interesting
nonlinear optical effects \cite{fedeli2018ultra,Lecz2018,Ong2021,Fedeli2018,
Curtis2021,Park2021}.

We propose an efficient ultra-high current accelerator composed of
nano-structures driven by arrays of ultra-short intense laser
pulses with potential applications in the context of ICF today and
fusion with fuels with large activation energies in the future.

We show that the fuel velocities the accelerator can generate can be
very high implying very high flow pressures. The accelerator is capable of
spreading fluid densities and velocities. Hence, fuels with activation
energies in the sub-$\text{MeV}$ range may come within reach.

High power short pulse laser technology has made impressive progress
over recent years \cite{strickland1985compression,danson2019petawatt}.
The total energy of the array of laser pulses is in the range of  $10
- 100 \, \text{kJ}$. Since the nano-structures in the accelerator can
be tailored to the optical laser radiation the coupling efficiency
between the lasers and the embedded accelerator is extremely high.

The numerical model used in the paper is capable of separating the
electronic time scales from the ones of the reactive ion flows
in each segment of the electro - ionic model. The electronic
parameters enter the ion flow equations as external parameters. Their
correct values can be determined by appropriate self-consistent
equations or parametrically depending on the computational load that
can be afforded. This way the code can simulate the fast electron
dominated accelerator and the reactive flow dynamics in the context of
kinetic reactive fluid dynamics.

The intent of the present paper is to introduce a new acceleration
concept for reactive fluids with a large range activation energies with the
help of ultra-fast, highly efficient electromagnetic acceleration of
ions. The acceleration process is based on nano-structures. The
nano-structures are irradiated by intense short pulse lasers.

The paper is structured as follows. After stating the numerical
model underlying our simulations in section \ref{model} in some detail
we explain the details of the accelerator concept in section
\ref{accel}. In section \ref{sum} we give a short summary and outlook.

\section{The model}\label{model}
Since we are investigating ultra-fast electrons and
fast high current reactive ionic flows in
the context of fusion the fully relativistic numerical model underlying our
investigations is kinetic and we base it on finite elements. While we
do not make use of the reactive flow part of the numerical model in
the present paper we quote some of its elements for future reference. The
analytical abstraction underlying the kinetic reactive flows is outlined to some extent in
\cite{ruhlkornarXiv}. The numerical model is an integrated adaptive
multi-scale model making use of tree data structures to manage
adaptive griding and elements of machine learning to support feed
forward error back integrators, which allows the simultaneous
investigation of the fast electronic as well as hydro time scales.

We obtain for $k=1, ..., N$, where $N$ is the number of finite
elements for the reactions
\begin{eqnarray}
\label{dnk}
\frac{dn_k}{dt} &\approx& - n_k \, \sum^N_{l=1}  n_l \, u_{kl} \, 
\sigma^{kl}_R \left( u_{kl} \right) \, \delta_{\vec r_k, \vec r_l} 
\end{eqnarray}
and the equations of motion
\begin{eqnarray}
  \label{drk}
\frac{d\vec r_k}{dt} &\approx& \vec u_k \, , \\
\label{dpk} 
\frac{d \vec p_k}{dt} 
&\approx& \frac{q_k}{V} \, \int_V d^3r \, \phi \left( \vec r, \vec r_k
\right) \, \left( \vec E_k + \vec u_k \times \vec B_k \right) \\
&&- \sum^N_{l=1} \nu_{kl}\left( u_{kl} \right) 
\, \left( \vec p_k - \vec p^{\, s}_{kl} \right) \, \delta_{\vec r_k,
   \vec r_l} \, . \nonumber
\end{eqnarray}
The equations and the symbols used are self-explanatory. The parameter
$\sigma^{kl}_R$ is the reactive cross section. The function $\phi$
denotes the form factor of a finite element. Each finite element has
the same density $n_0$. The remaining parameters to augment the force
equations are
\begin{eqnarray}
\label{nukl}
\nu_{kl} \left( u_{kl} \right) 
&\approx& \frac{n_l}{C_{kl} \, \gamma^2_{kl} \, 
u^3_{kl}} \, \Lambda_{kl} \, , \\
\label{lamkl}
\Lambda_{kl}
&\approx&\ln  \left( \frac{u^2_{kl}+B}{B}  \right) -
\frac{u^2_{kl}}{u^2_{kl}+B} \, , \\
\label{Ckl}
C_{kl} &=&\frac{4 \pi \epsilon^2_0 m^2_{kl}}{q^2_k q^2_l} \, , \\
\label{mkl}
m_{kl}
&=&\frac{m_k \, m_l}{m_k + m_l} \, .
\end{eqnarray}
The parameter $B$ is an effective shielding parameter. The remaining
kinematic relations are
\begin{eqnarray}
\label{uk}
\vec u_k&=&\frac{c\vec p_k}{\sqrt{m^2_kc^2 + \vec p^{\, 2}_k}} \, , \\
\label{ukl}
u_{kl}&=&\frac{c \sqrt{\left( p_k \cdot p_l \right)^2 - m^2_km^2_l  
                c^4}}{p^{0i}_k p^{0j}_l} \, , \\
\label{pkls}
\vec p^{\, s}_{kl} &=& \gamma_{kl} \, \vec{\beta}_{kl} \, p^0_{k}  -
\frac{1}{\beta^2_{kl}} \, (\gamma_{kl} -1) \, \left( \vec{p}_{k} \cdot 
\vec{\beta}_{kl} \right) \, \vec{\beta}_{kl} \, , \\
\label{betakl}
\vec{\beta}_{kl} &=& \frac{\vec{p}_k+\vec{p}_l}{p^0_{k}+p^0_{l}} \, , \\
\label{gamkl}
\gamma_{kl} &=& \frac{1}{\sqrt{1-\beta^2_{kl}}} \, ,
\end{eqnarray}
where care has to be taken if one of the colliding constituents is an
electron. For the sake of brevity we do not state details.

Equations (\ref{dnk}) - (\ref{mkl}) are obtained by decomposing
the reactive kinetic equations into reactive resistive motion and
kinetic lateral broadening of the binary correlation operators. We do
not quote the lateral broadening part of the
model here. We note that lateral broading is qute important for
energy - momentum conservation. Neglecting it underestimates
range, a typical parameter in reactive flows. In addition, to include
radiative processes we have to add radiation reactive terms on the
mean field level of the model. On the binary correlation level the
binary transition amplitudes have to be generalized to the radiative
one. Equations (\ref{dnk}) - (\ref{mkl}) are augmented with Maxwell's
equations.

\section{High efficiency ion acceleration}\label{accel}
A concern is the efficiency of the indirect drive ICF approach to
fusion. We propose an embedded accelerator technology for fluids that
makes use of embedded nano-structured accelerator configurations for
the acceleration, convergence, and if required the divergence of ionic
flows at very high efficiency and precision.

The configuration is powered by an array of efficient short-pulse lasers
with moderate individual pulse energies in the range of $100 \,
\text{J}$. They can generate ion kinetic pressures and convergence 
velocities that are orders of magnitude higher than what is state of
the art at present.

The coupling between the laser pulses and
appropriately tailored nano-rods can be extremely efficient. It
is important to note in the context that the observed efficient laser
- rod coupling is only possible because advanced laser pulse
technology exists. The laser pulses this technology provides are
capable of propagating into tailored configurations of nano-rods at
the speed of light outrunning the impact of their interaction with the
rods. Hence, the nano-structures cannot shield themselves
from the absorption of the related extreme energy densities. By
combining nano-structures and the proposed laser arrays a high degree
of spatio-temporal control over the deployment of extreme optical energy
densities on the fastest possible time scales in the accelerator at high
efficiency levels can be expected.

\begin{figure}[ht]
\begin{center}
\includegraphics[width=20mm]{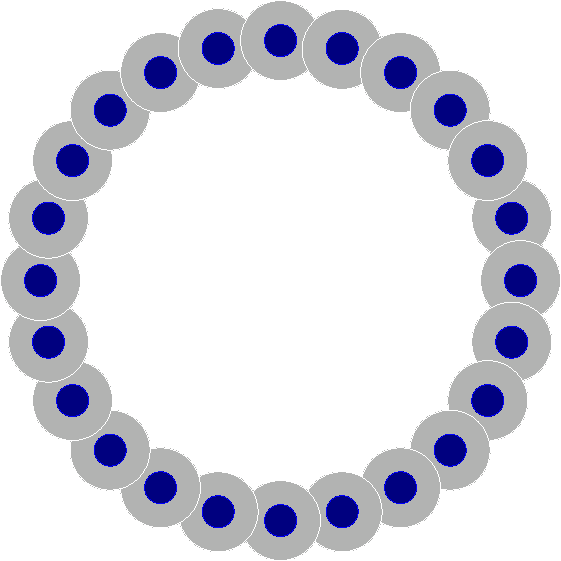}\\
\vspace{0.5cm}  
\includegraphics[width=20mm]{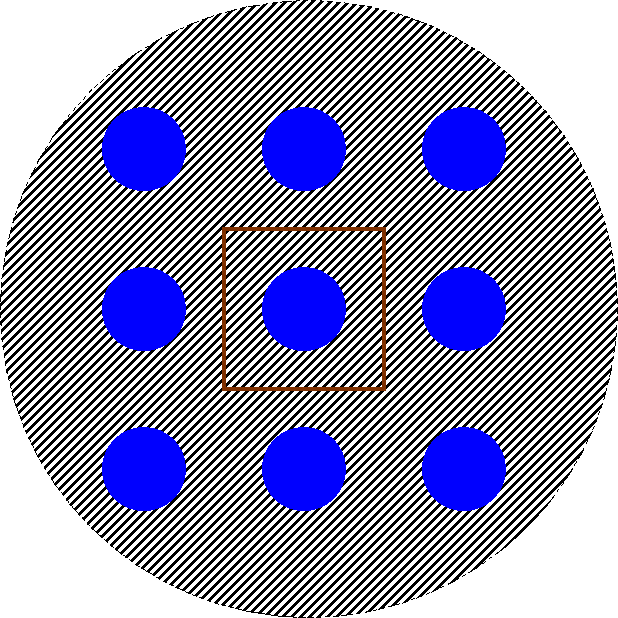}\\
\vspace{0.5cm}  
\includegraphics[width=20mm]{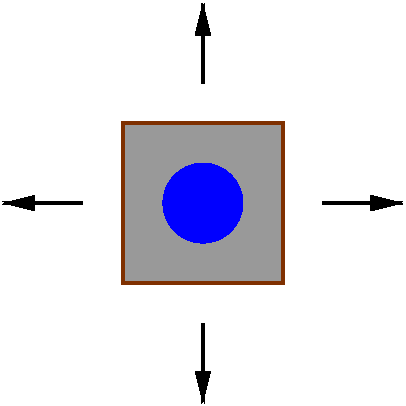}\\
\end{center}
\caption{\label{irrad_rods} Sketch the hypothetical 2D accelerator and the 
  irradiation pattern by the proposed multi-lasers (plot at the top), the 
  irradiated elements of the accelerator (2nd from top), and the 
  simulation box for the laser - rod interaction (bottom plot). The 
  nano-rods are depicted in blue, the multi-lasers in shades of gray, and 
  the simulation box boundaries in brown. To reduce computational load 
  we take a single rod and periodic lateral boundaries in an effort to 
  mimic the mutual correlations between the rods. Each laser pulse 
  irradiates between $10^8 - 10^9$ nano-rods depending of their 
  setup. The accelerator contains up to $10^{10}$
  nano-structures. For the nano-geometry of the rods depicted and the 
  morphology simulated the ion beam is launched in radial 
  direction. However, the launch direction of individual fluids is a 
  free parameter. The center of mass of the accelerator must remain at rest.}
\end{figure}

For reasons of simplicity we stick to a cylindrical macro accelerator
geometry in this paper, while the nano-rod simulations are fully 3D.
The 2D accelerator layout is sketched in Fig. \ref{irrad_rods}. Also a
hypothetical 3D accelerator layout is sketched in the figure. The
accelerator is comprised of multiple segments of nano-structured
elements colored in blue in the figure. The nano-structured segments
are irradiated by multiple laser pulses as
depicted in the figure. Each laser pulse is capable of irradiating
between $10^8 - 10^{9}$ nano-structures. Each single irradiated
nano-rod is capable of releasing up to $10^{10}$ accelerated ions. The
macro accelerator is capable of generating currents between $10^{18} -
10^{19}$ ions in about $500 \, \text{fs}$. The number of individual 
lasers in the array is about $1000$. The required energies of the ions range
between $0.1 - 2.0 \, \text{MeV}$.

The expected ionic implosion velocities and flow 
pressures are orders of magnitude larger than those achievable with 
thermal ablation. There are conceptual similarities with the direct 
drive ICF approach \cite{craxton2015direct}. Only the energy densities 
and spatio-temporal parameters achievable in the realm of fusion with 
advanced laser technology are orders of magnitude larger. 

\begin{figure}[ht]
\begin{center}
\includegraphics[width=60mm]{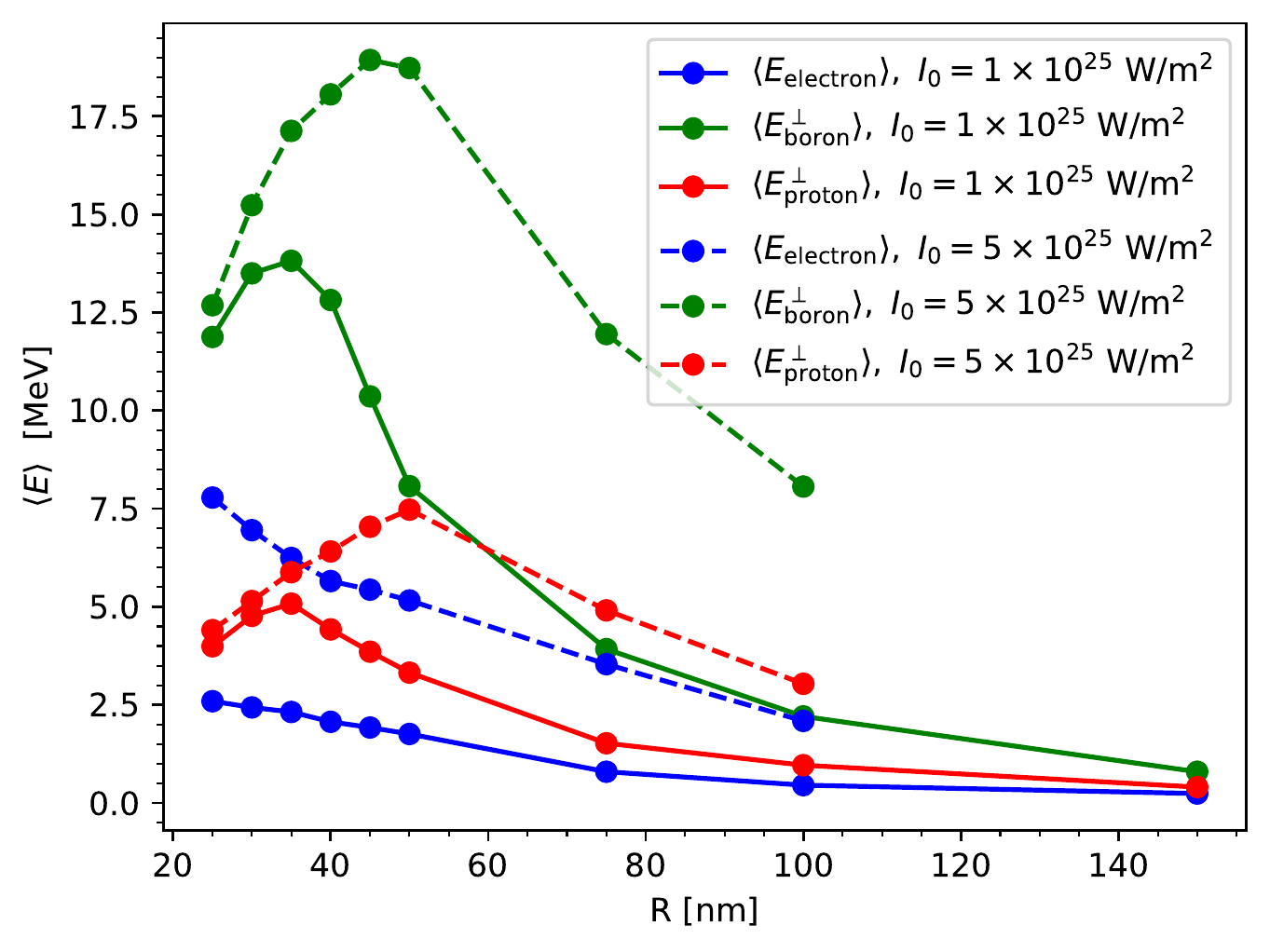}\\  
\end{center}
\caption{\label{rod_explosion} Illustration of the explosion of a single 
nano-rod exposed to an ultra-short high-field laser pulse with up to 
$30 \, \text{fs}$ FWHM at $\lambda = 400 \, \text{nm}$. The figure 
shows the energy of the individual ions. Their energies can be 
tailored towards the peak of the fusion cross sections. The $x$-axis 
shows the radii of the simulated rods. The $y$-axis gives the energy 
per charged particle. The rods consist of a heavy substrate and a 
lighter reactive fluid. The substrate in the simulation is 
$\ce{^{11}B}$ and the light fluid consists of protons. By reducing the 
radii of the rods, their morphology, and the laser intensity 
less energetic ions can be generated. To emulate the 
interaction with multiple nano-rods periodic boundaries have been 
chosen as detailed in the text.}
\end{figure}

In Fig. \ref{rod_explosion} the ion energies obtained from a single
nano-rod with varying radii interacting with short laser pulses up
to $30 \, \text{fs}$ FWHM at $\lambda=400 \, \text{nm}$ at the intensities of
$I_<\approx 1.0 \cdot 10^{21} \, \text{Wcm}^{-2}$ and $I_>\approx 5.0 \cdot
10^{21} \, \text{Wcm}^{-2}$ are shown. The simulated rod contains
the same uniform number densities $n_p = n_B = 10^{23} \,
\text{cm}^{-3}$ for protons and boron. The simulated rod is $16 \, \mu
\text{m}$ long, while the simulation box is $32 \, \mu \text{m}$
long. The lateral boundery conditions are periodic, while front and
back sides of the box are transparent. The numerical resolution is
$\Delta x = \Delta y = \Delta z = 5.0 \, \text{nm}$. The temporal
resolution is about $\Delta t \approx 7.2 \, \text{as}$.

A prominent feature in Fig. \ref{rod_explosion} is the convergence
of the single particle energies of the protons and the boron ions to the
same values for the two intensities $I_<$ and $I_>$ for rod radii
$R<40 \, \text{nm}$. The process underlying the acceleration
of the protons and the boron ions for the rod radii $R<40 \,
\text{nm}$ are Coulomb explosions. For large rod radii the conversion
efficiency of the laser energy to the ions is small as expected since
large rod radii exceed the skin depth by far. Since the energies of
the protons and the boron ions are about the same for both intensities
the energy conversion efficiency into ion energy increases more than
$5$-fold for the lower laser intensity since also the energy deposited
into the electrons drops substantially at the lower intensity
$I_<$. The implication is that nano-rods can be engineered to
accommodate the desired nonlinear optical properties for the driver
lasers and to transfer as much energy to the ions as possible while
limiting electron energies. In Fig. \ref{rod_explosion} it is shown that
there is an energy gap between the protons and the boron ions. The
boron ions are about $3$-fold as energetic as the protons. This
observation can be understood taking the charged fluid expansion into
account. The theoretically predicted factor is about $3.5$ due to the
particular shape, size and composition of the rods. In important conclusion is
that the energy gap between fluids from rod with multi ionic
constituents can be engineered.

\begin{figure}[ht]
\begin{center}
\includegraphics[width=60mm]{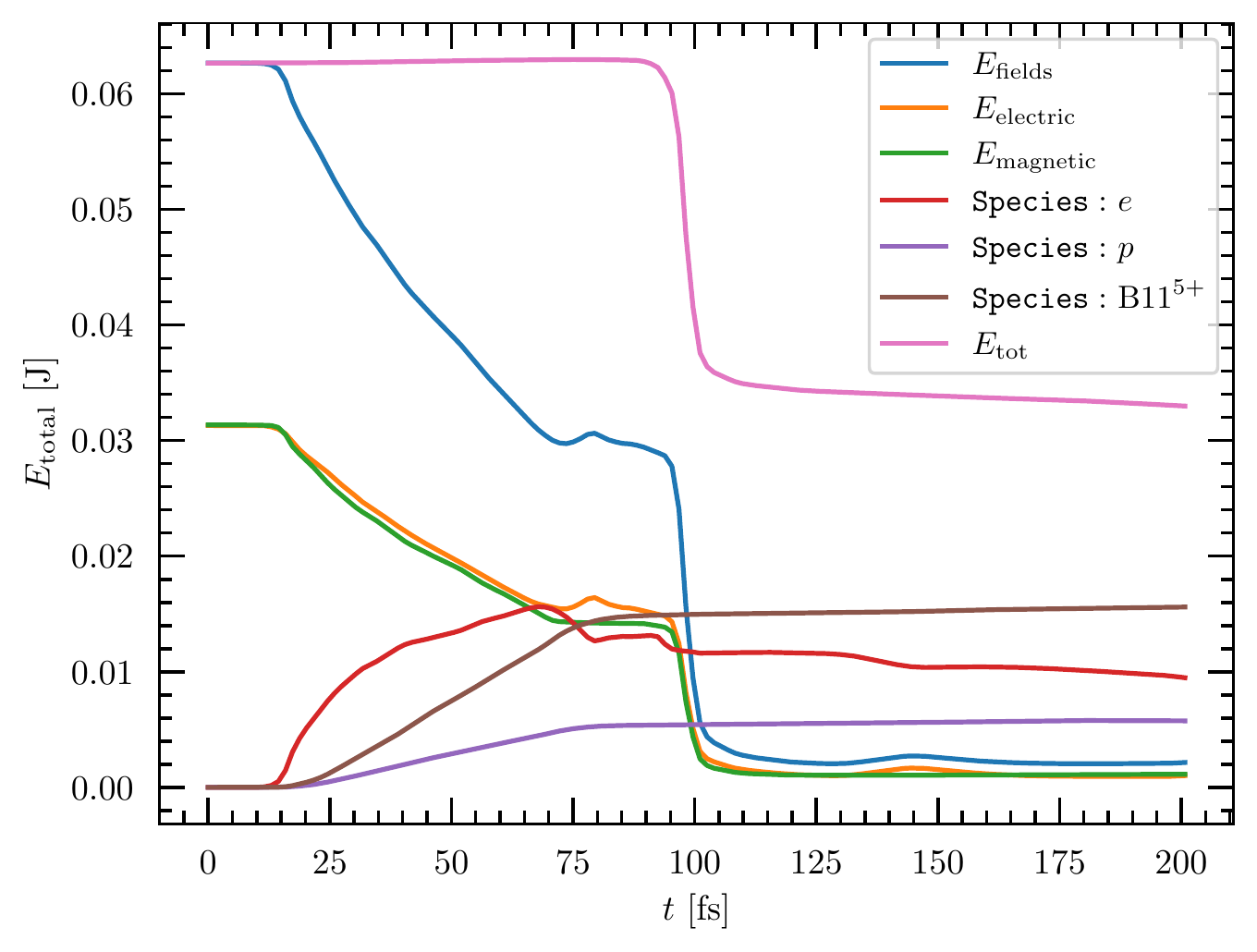}\\  
\end{center}
\caption{\label{rod_efficiency} Illustration of the energy balance in 
the simulation box. Shown are the electromagnetic field energies, the 
electron energy, and the proton and boron energies as functions of 
time. As is seen both the proton and boron energies grow linearly with 
time and hence laser propagation depth. A second implication 
is that the nonlinear optical properties of the system seem to be 
under control. A third feature is the energy conversion efficiency 
into ions, which according to the figure is $>60 \, \%$. Almost
the total absorbed laser energy is transformed into ionic motion. A
total of about $10^{10}$ ions are accelerated. The simulation
parameters are detailed in the text.}
\end{figure}

In Fig. \ref{rod_efficiency} the energy balance in the simulation box
as a function of the interaction time is monitored. The figure
illustrates that there is a linear rise of the total ion energy for
both the protons and the boron ions starting at about $t= 25 \, \text{fs}$
until at about $t=75 \, \text{fs}$. This corresponds to the time the
impinging laser pulse needs to travel a length of $16 \, \mu\text{m}$
at speed of light. At $t=25 \, \text{fs}$ the laser pulse hits the
nano-rod and at $t=75 \, \text{fs}$ it leaves the rod. At $t=100 \,
\text{fs}$ the laser pulse exits the simulation box, which implies
that the electromagnetic energy density in the simulation box must
drop rapidly. The total energy in the simulation box is now in the
ions and electrons with only little field energy left. The implication
is that the system of charged particles is quasi-neutral at this
point. The simulation is stopped at $t = 200 \, \text{fs}$. It is
apparent from the figure that the ions retain their kinetic energies
while the electrons loose some of theirs since some of them exit the
transparent front and back faces of the simulation box due to the
specific numerical boundaries set up there. From
Fig. \ref{rod_efficiency} we conclude that $>60 \, \%$ of the
deposited laser energy has been converted to the ions. We note that
similar efficiency results should hold for somewhat longer laser
pulses since the rods can be tailored towards the incident laser radiation.

\begin{figure}[ht]
\begin{center}
\includegraphics[width=40mm]{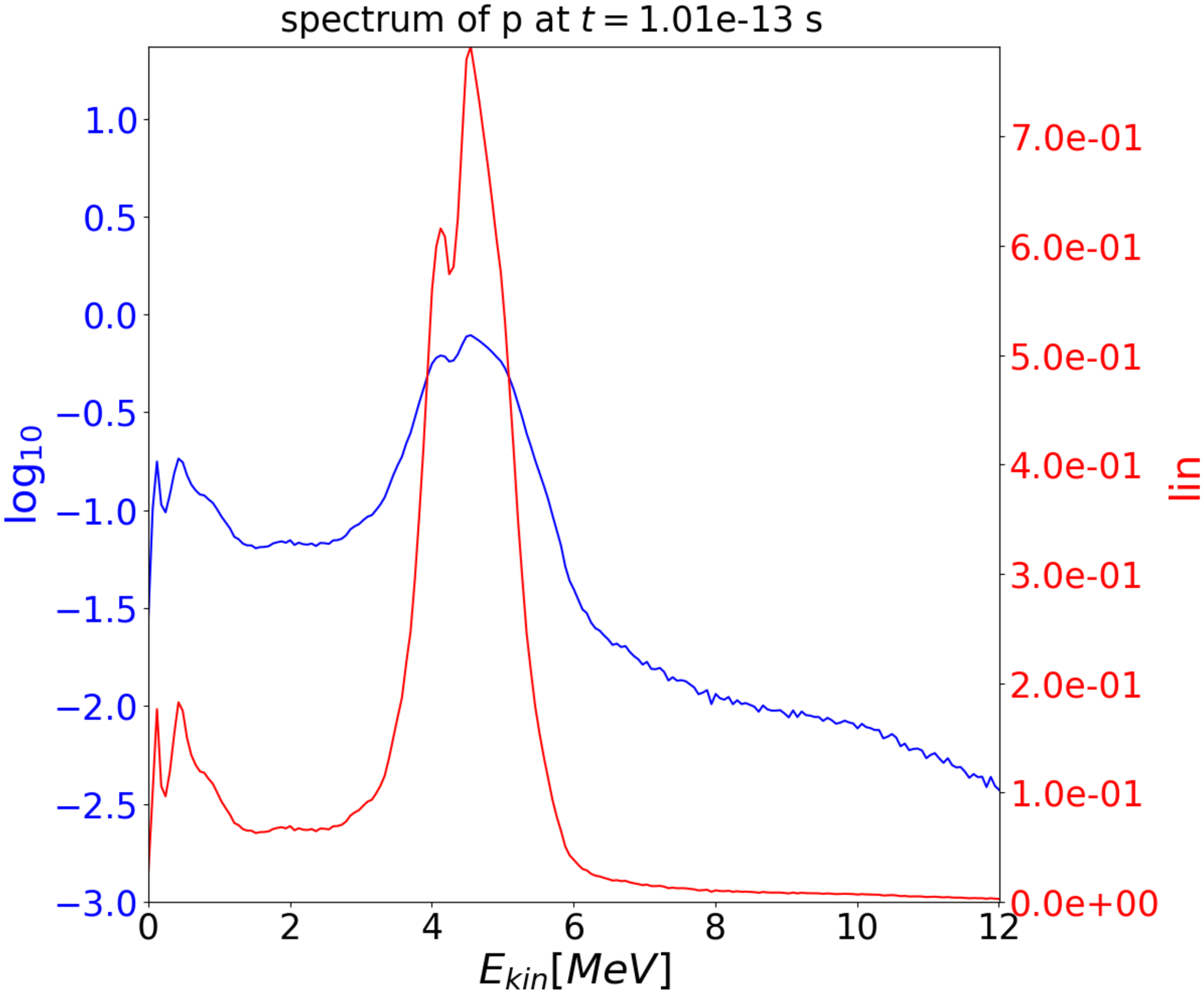}
\includegraphics[width=40mm]{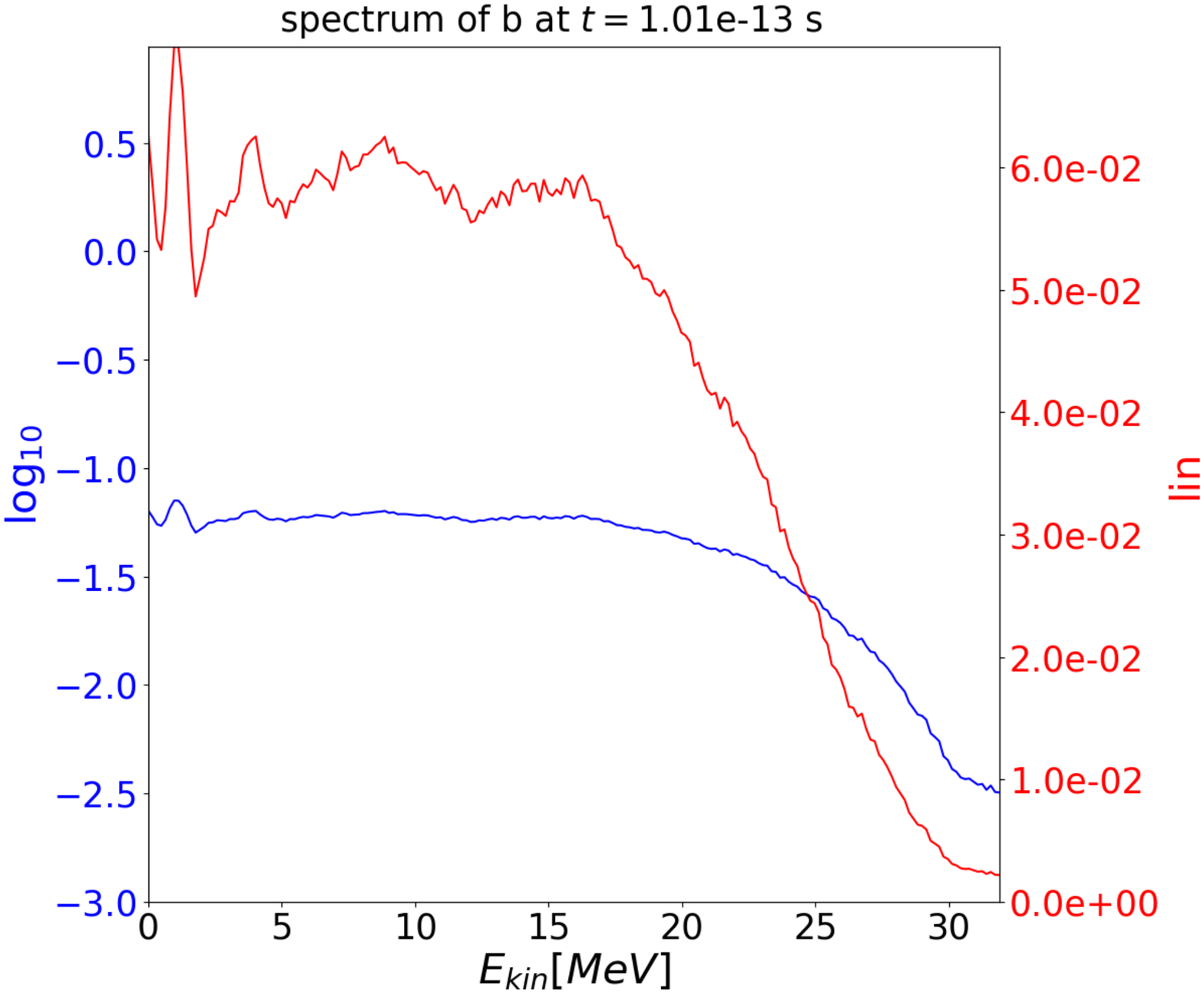}    
\end{center}
\caption{\label{rod_spectrum} Energy spectra for protons and boron
ions generated by a rod with $R=25 \, \text{nm}$, which is the small
radius regime. The upper plot shows the protons and the lower plot
the boron ions. The energy of the protons is sharply
peaked. By changing the size and morphology of the nano-structures the
ion spectra can be tailored, specifically the boron ion spectra can
become much sharper. The simulation parameters are the same as in
Fig. \ref{rod_explosion}.}
\end{figure}

Figure \ref{rod_spectrum} shows the simulated proton and borons
spectra for a rod with $R=25 \, \text{nm}$ and for the intensity
$I_<$. While the spectrum of the protons is sharply peaked at about $4
\, \text{MeV}$ the boron ion spectrum is broad. This leads to the
separation of proton and boron fluids preventing thermalization
between the species during rapid convergence. Making use of lower
laser intensities and optimized rod morphologies the energy convergence
efficiency and spectral signatures of both fluids can be considerably
improved further.

\begin{figure}[ht]
\begin{center}
  \includegraphics[width=20mm]{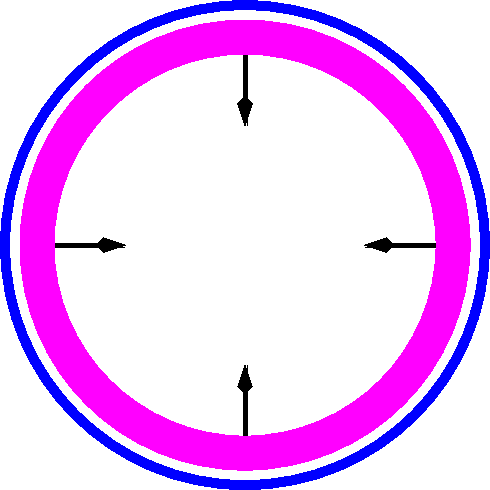}\\
  \vspace{0.5cm}
  \includegraphics[width=30mm]{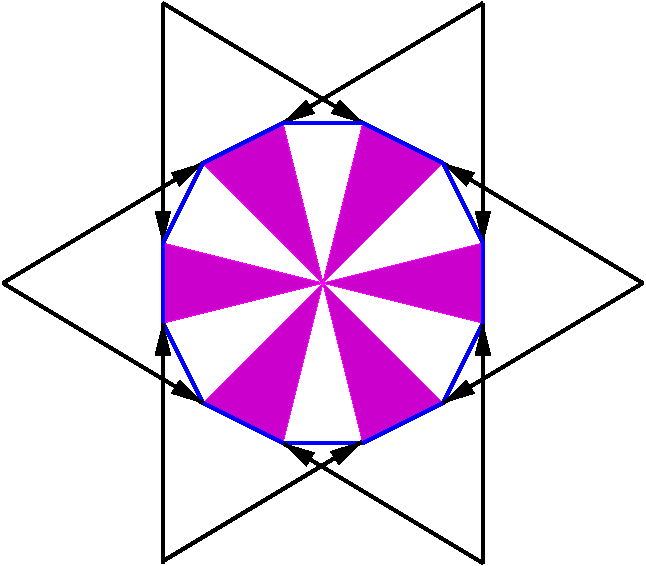}\\    
\end{center}
\caption{\label{imploder} Top view of a hypothetical cylindrical 
onion shell accelerator - converter layout (top plot). The converter 
fluid colored in pink is driven by the accelerator colored in blue. The 
accelerator is capable of generating large ionic currents.  A side 
view on a principal 3D onion shell accelerator - converter layout with 
is rotational symmetry about the $x$-axis is shown in the lower plot. 
The black arrows illustrate the directions of impinging individual 
laser pulses.}
\end{figure}

Sketches of a few principal accelerator configurations intended to
illustrate the flexibility of the concept are depicted in
Fig. \ref{imploder}. The figure shows a 2D cylindrical layout and a
possible 3D configuration. The accelerator is capable of generating ion
flows in the range of $10^{19} - 10^{20}$ ions with a large range of
tunable energies.. The ion energies obtained can be engineered to a
large extent. The flows can be generated in much less
than a $1 \, \text{ps}$. The energy contained in the accelerated ions
can range between $10^{23} - 10^{24} \, \text{eV}$ due to the high
conversion efficiency that can range up to  $80\%$. 

\section{Conclusions}\label{sum}
Intense ultra-short laser pulses interacting with tailored
nano-structures are capable of generating converging ion flows with
$10^{20}$ ions in a large range of tunable energies on sub-ps time scales.

The properties of these ionic flows may potentially be applied to the
design of future reactors for energy production with fuels with
sub-MeV activation energies and the formation of hot spots for fusion
energy production in the context of isentropic compression schemes.

Specifically, the numerical and experimental investigation of fast
reactive flows for fusion energy production in the scope of high
activation energy aneutronic fuels like $\ce{p^{11}B}$ is our primary goal.

\section{Acknowledgements}
The present work has been motivated and funded by Marvel Fusion
GmbH. We would like to thank the Marvel Fusion team and in particular
Christian Bild, Matthias Lienert, Markus Nöth, Gaurav Raj, Michael
Touati, and Naveen Yadav.

\bibliographystyle{elsarticle-num} 
\bibliography{literatur_eqn_motion}

\end{document}